# Supervised learning applied to high-dimensional millimeter wave transient absorption data for age prediction of perovskite thin-film

Biswadev Roy*, A. Karoui, B. Vlahovic, and M.H. Wu
Department of Mathematics and Physics,
North Carolina Central University,
1146, Mary Townes Science Complex, 1900 Concord Street, Durham, NC 27707, U.S.A.
*Address for communication: broy@nccu.edu

**Abstract**

We have analyzed a limited sample set of 120 GHz, and 150 GHz time-resolved millimeter wave (mmW) photoconductive decay (mmPCD) signals of 300 nm thick air-stable encapsulated perovskite film (methyl-ammonium lead halide) excited using a pulsed 532-nm laser with fluence 10.6 µJ cm$^{-2}$. We correlated 12 parameters derived directly from acquired mmPCD kinetic-trace data and its step-response, each with the sample-age based on the date of the experiment. Five parameters with a high negative correlation with sample age were finally selected as predictors in the Gaussian Process Regression (GPR) machine learning model for prediction of the age of the sample. The effects of aging (between 0 and 40,000 hours after film production) are quantified mainly in terms of a shift in peak voltage, the response ratio (conductance parameter), loss-compensated transmission coefficient, and the radiofrequency (RF) area of the transient itself (flux). Changes in the other step-response parameters and the decay length of the aging transients are also shown. The GPR model is found to work well for a forward prediction of the age of the sample using this method. It is noted that the Matern-5/2 GPR kernel for supervised learning provides the best realistic solution for age prediction with $R^2 \approx 0.97$.

## 1.0 Introduction

Direct band-gap hybrid metal-organic perovskite exhibits balanced carrier mobility [1], considerably long electron-hole recombination length, and low non-radiative Auger recombination, making it a very ideal material for harnessing solar energy, as radiation detectors [2], as THz sensing, modulation, and imaging [3]. This material exhibits excellent charge carrier mobility and recombination parameters [4]. Lifetime is affected by 4 principal mechanisms [5] such as recombination via trap states, polaron formation (an increase of the electron effective mass), Rashba effect, and photon recycling. However, polaron dynamics and photon recycling affect the charge-carrier recombination lifetime to a greater extent than the former two processes [5, 6]. Kinetic and frequency analysis of the perovskite sample using the time-resolved microwave conductivity (TRMC) method is found to show significant suppression of recombination with extremely low concentration (<10%) and very shallow trap-depth (10 meV) [7]. Methyl ammonium lead halide (MAL3I) thin films also exhibit grain size limited mobility i.e. the surface grains represent represents the true electronic domain size and the time scale of diffusive transport across grain boundaries is slower than the microwave field when probe frequency is 8.9 GHz [8]. In addition, the low cost and versatility of its fabrication enhance the application potential of this material for solar energy conversion. This film also exhibit high absorption efficiency in the 300-900 nm spectral range, which is higher than the absorption of silicon in the UV-blue spectrum.

There is a good amount of other 3D MAL3I that is proven by its excellent characteristics as a photovoltaic material with efficiencies crossing 25%, is applicable for fabrication of light-emitting diode (LED), photodetection, biochemical applications, radiation detection, scintillators, and is an excellent THz switch material since it exhibits carrier-induced dielectric property changes [9]. The sensing application is pronounced because of its exceptional hydration-dehydration, electronic and phase transition, and reversible ion-insertion capability which makes it very attractive [10]. MAL3I-added PVDF (polyvinylidene fluoride) nanofibers exhibit piezoelectric properties and can be used as piezo-/pyroelectric nanogenerators [11].

We proposed to study the MAL3I property in mmW domain by using the same encapsulated MAL3I sample for the time range of 0 to 40,000 hours elapsed after the film production. The study of age dependence of carrier response and dynamics will offer an opportunity to merge structure inhomogeneity (due to the aging process) and correlated measurements of charge carrier density, dynamics, and mobility obtained at millimeter wave frequencies. Baranov et al. (2021) [12] have published reports of enhanced photoluminescence in self-assembled lead halide perovskite nanocrystal superlattices resulting from the contraction of superlattices and coalescence of nanocrystals over time. In our experiments, we have studied the consequences of changes in photoconductance over aging periods of the sample.

TRMC is a pump-probe spectroscopy technique [13] and is an excellent method of capturing the charge carrier dynamics [14] especially, if the charge carrier density and the photon-to-carrier conversion efficiency can be quantized then using the TRMC electric field amplitude response $\Delta V/V_0 = -K\Delta P/P_0$ =KCF, $\Delta V$, and $\Delta P$ being the peak voltage and peak power registered by the detector when the laser is turned on, K is the calibration constant, C is the response-laser fluence slope and F is laser fluence itself. Using the voltage response from detector, it is very easy to ascertain the mobility of the carriers directly. We developed a TRMC-like apparatus in-house and named time-resolved millimeter wave conductance apparatus (TR-mmWC) [13]. This system operates in the 110 GHz-178 GHz, and employs a completely free space quasi optical (Gaussian beam) probe beam through the sample. The system can be used for transmission and reflection mode, in these age related property experiments we have used transmission mode only. Photoconductive decay (mmPCD) for studying recombination characteristics are very efficiently acquired using a high sampling rate digital storage oscillioscope (20 GSa/s). Passage of millimeter wave probe signal through the pump-excited sample is subjected to the local changes of the dielectric response of the medium, and as such the excited-state absorption occurs due to attenuation of the probe photons.

Machine learning techniques have been applied largely to study the stability and predictability of the fate of solid-state materials [14, 15]. Users of electronic devices seldom do not carry ready information about the age of the associated components, hence, there is a need to infer age of the sample using the mmW signal responses. This would be of great advantage for inferring the life cycle of the operating device and to ascertain its performance. Machine learning can play a very effective role in sample age assessment especially if the properties of the material used in the components are very difficult to measure and are expensive, or, there are non-deterministic data merging out of the tests due to the complexity of the phenomenon being observed with the device having aged material [16]. Physics-driven machine learning has been applied to infra-red data for characterizing surface microstructures of complex materials for better material performance [17]. The importance of computational materials design has been highlighted especially for the design of crystalline solids in which application of first-principle theory such as density-functional theory-based calculations for single and binary solids [18], and related simulation becomes overwhelmingly complex and could potentially be handled only by researchers and not operationally [19].

The present paper focuses on studying the age-dependent charge carrier recombination processes as a function of age in hours elapsed after perovskite film production. We inter-compare the aging mmW responses using the acquired averaged signal characteristics such as the departures in peak AC voltage, the TRmmWC conductance parameter $\Delta V/V_0$, the signal transmission coefficient (ratio of the through sample RF voltage at the detector to the free space voltage at the same detector), and the radio-frequency area (integral of the voltage-time product in positive area of transient, $\Delta V(t)$). We study 8 signal-dependent step-responses (directly from the averaged transients collected at various ages of the MAL3I sample) and 4 derived charge carrier dynamical parameters from the TR-mmWC datasets obtained throughout 0 to 40,000 hours.

We select 5 features out of the 12 signal parameters that bear strongest negative correlation with the sample age in hours to develop a Gaussian Process Regression (GPR) machine-learning model [20, 21]. We have used the same set of dataset of the sample for training and testing the age-prediction algorithm. Since the number of signal features used (12) and the number of observations is comparable to 11 days we term the data to be "high-dimensional". Zhang et al., (2018) [22] have used high dimensional data with GPR kernels to train the banknote data from the UCI machine learning (ML) repository. They have reported that GPR models work very well for limited datasets, and predictions are satisfactorily comparable to other established modeling techniques such as extended nearest neighbor (ENN), the k-nearest neighbor (KNN), naïve Bayes, linear discriminant analyses (LDA), and the multilayer perceptron (MLP) techniques. Hence, we adopt the GPR technique using the mmW responses over a long period, and intercompare the prediction accuracy obtained after using different kernels with or without cross-validation, and report its responsiveness.

**2.0 Perovskite sample and the experiment**

Glass encapsulated Methylammonium lead Iodide (MAPbI$_3$) thin film on a quartz substrate is obtained from North Carolina State University (produced on May 02, 2017) [23, 24]. The sample has been characterized using millimeter wave (mmW) transient at probe frequencies of 120 GHz and 150 GHz. The 120 GHz probe wave was generated from the cavity IMPATT diode oscillator (CIDO) and 150 GHz was generated from the backward wave oscillator (BWO) source respectively. Simultaneously, in our time-resolved TR-mmWC apparatus, the glass-covered MAL3I sample was excited by a 1kHz triggered 532-nm laser beam (~20 µJ/pulse) that has a width of 0.69 ns. The maximum fluence of 10.6 µJ/cm² and a laser spot size of 10 mm was applied to the sample. The ratio of laser to probe beam spot sizes is almost 4:1, and the sample was held static to the vertically polarized beams (both, the pump and probe).

**Table 1** Provides the dates, the DC (probe signal through the sample with no laser excitation), peak radiofrequency (RF) voltages, and decay length durations at the dates of each experiment in sequence.
*Experiment ID 19 is included in the transient behavioral analysis but excluded from the exponential fitting and related analyses.
**These data are used only for age prediction using transient signal responses employing the supervised regression learning

| ID number | 150GHz, Experiment Dates | Age (Hours) | DC (mV) | Peak Positive RF Voltage (V) | Peak Negative RF Voltage (V) | Positive-decay length (s) |
|---|---|---|---|---|---|---|
| 11 | 05-03-2017 | 0 | 327 | 45 x 10$^{-4}$ | -2.03 x 10$^{-5}$ | 2.5 x 10$^{-7}$ |
| 02 | 12-18-2017 | 5,496 | 168 | 37 x 10$^{-4}$ | -1.1 x 10$^{-4}$ | 1.3 x 10$^{-5}$ |
| 09 | 01-31-2018 | 6,552 | 260 | 37 x 10$^{-4}$ | -1.1 x 10$^{-4}$ | 1.3 x 10$^{-5}$ |
| 03 | 02-21-2018 | 7,056 | 290 | 37 x 10$^{-4}$ | -2.8 x 10$^{-5}$ | 1.8 x 10$^{-7}$ |
| 04 | 05-03-2018 | 8,760 | 189 | 23 x 10$^{-4}$ | -8.3 x 10$^{-6}$ | 7.3 x 10$^{-7}$ |
| | **120GHz, Experiment Dates** | | | | | |
| 08 | 10-20-2020 | 30,408 | 194.5 | 4.7 x 10$^{-5}$ | -5.8 x 10$^{-6}$ | 1 x 10$^{-6}$ |
| 07 | 03-08-2021 | 33,720 | 216 | 1.3 x 10$^{-4}$ | -3.8 x 10$^{-6}$ | 1.6 x 10$^{-6}$ |
| 13 | 06-03-2021 | 35,808 | 119 | 1.2 x 10$^{-4}$ | -2.1 x 10$^{-6}$ | 1.6 x 10$^{-6}$ |
| 19* | 08-23-2021 | 37,752 | 50.7 | 5.8 x 10$^{-5}$ | -2.6 x 10$^{-6}$ | 2.8 x 10$^{-6}$ |
| 23** | 10-01-2021 | 38,688 | 100.3 | 6.7 x 10$^{-5}$ | -2.8 x 10$^{-6}$ | 1.6 x 10$^{-6}$ |
| 25** | 11-09-2021 | 39,624 | 70.5 | 3.1 x 10$^{-5}$ | -1.6 x 10$^{-6}$ | 1.6 x 10$^{-6}$ |

Transient (RF) and DC transmission data of probe signal (referred to as mmPCD) through sample were collected using a 6 GHz input bandwidth digitizer after amplifying Schottky detector (RF) signal. Signal averaging was done 4,096 times using the 20 GSa/s digitizers, and data were collected on 11 different dates spanning almost 40,000 hours from the time of manufacture. The peak voltage through the sample remains consistently within the 3.5 to 3.6 mV range. Table I explains the aging sample experimental dates, measured properties, and sequence of experiments performed at 150 and 120 GHz respectively.

The 150 GHz system operated 5 times between 05-03-2017 and 05-03-2018 and thereafter, the BWO tube was expended resulting in changing the source of the probe signal to an IMPATT oscillator with a center frequency of 120 GHz. Although, the signal power level change between 120 and 150 GHz is only -0.032 mW, and the differences in material responses to these changed probe frequencies are not very significant, hence, we believe, we could analyze the complete aging behavior by considering data obtained at both these probe frequencies, hence, we have combined the data to create more consistent sample space. Although, the 120 GHz probe signals will form the basis of the analysis as it is one of the existing operable sources in our possession, and test data are being routinely generated as the MAL3I sample is still in working condition.

## 3.0 Results
### 3.1 Photo-response signal and its characteristics by age

The raw mmPCD datasets obtained using TRmmWC at 150 and 120 GHz (by sample age in hours elapsed from the film production date) have been shown in Figure 1. The 150 and 120 GHz photo-response (with laser on) peak voltages ($\Delta V$), The millimeter wave through sample voltage ($V_{dc}$) registered by the Schottky barrier zero-bias diode (ZBD) [25] when the laser is turned off, and the the decay length (time in second between the onset and rise to peak and decay to a zero voltage) of all the transients are shown in figs. 1(b) and 1(c) respectively. Figure 1(a) shows the $\Delta V$ profiles (in log scale) for the aging perovskite at 150 GHz at ages between 0 and 8760 hours and 120 GHz at sample ages between 30,000 and 40,000 hours respectively. Although, the averaged transients in Fig 1(a) show the peak voltages not steadily declining with age, at some particular ages it is showing an upward trend and at some, there is a steady decline. This is because the peak voltages are obtained at slightly different probe power at each experiment, and the peak voltages need to be normalized by the dc through sample voltages when not excited by the laser. It will be seen that the TR-mmWC response ratio ($\Delta V/V_{dc}$) shows a steady decline with age as shown in Fig. 2(c) appearing in the next sub-section. However, figures 1(b) and (c) are two discrete plots showing the RF peak voltages ($\Delta V$), the DC voltage through the sample, $V_{dc}$ (for fixed beam powers at 120, and 150 GHz), and the complete recombination period (the complete transient decay length in seconds) as a function of sample age for 150 GHz and 120 GHz respectively. It is noted, in general, from both figures 1(b) and 1(c) that there is a steady decrease in $\Delta V$ and DC voltage with time in both probe frequencies. At 150 GHz, the changes in recombination time that are recorded at various sample ages might depict the photon recycling activity [5] and may be attributed to the time variation of radiative efficiency of perovskite MAL3I. It shows an increase in magnitude until about 5,000 hours to 6,500 hours (evolution from 100s of ns to microseconds) and after which, it fell off to 100 ns after 7,000 hours of aging. A similar trend is also seen at 120 GHz probe frequency between 35,000 and 40,000 hours elapsed time (c.f. Figs 1(b) and 1(c) $V_{dc}$ trends).

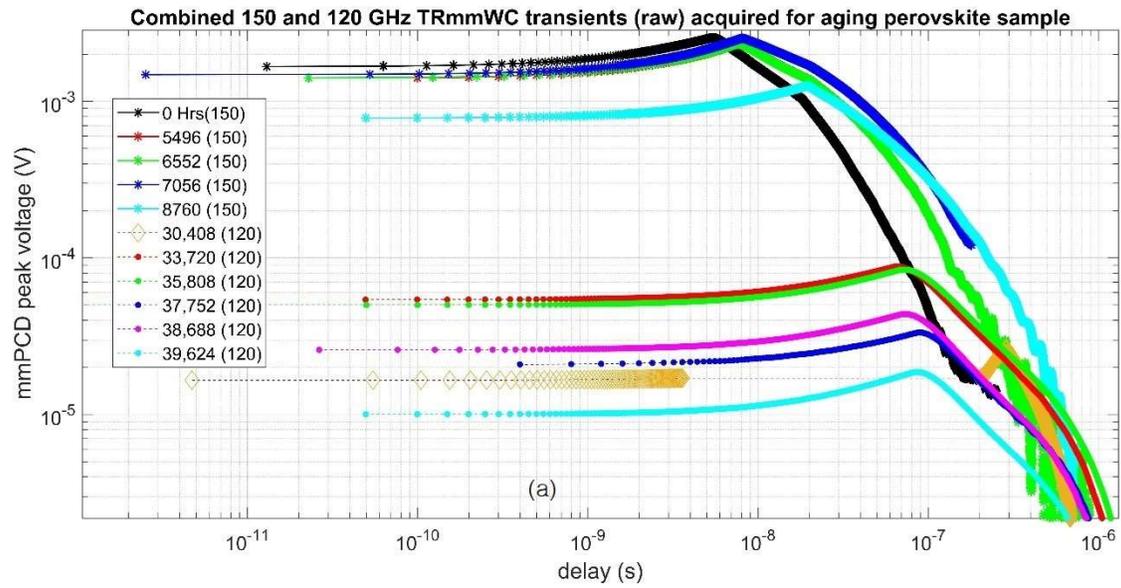

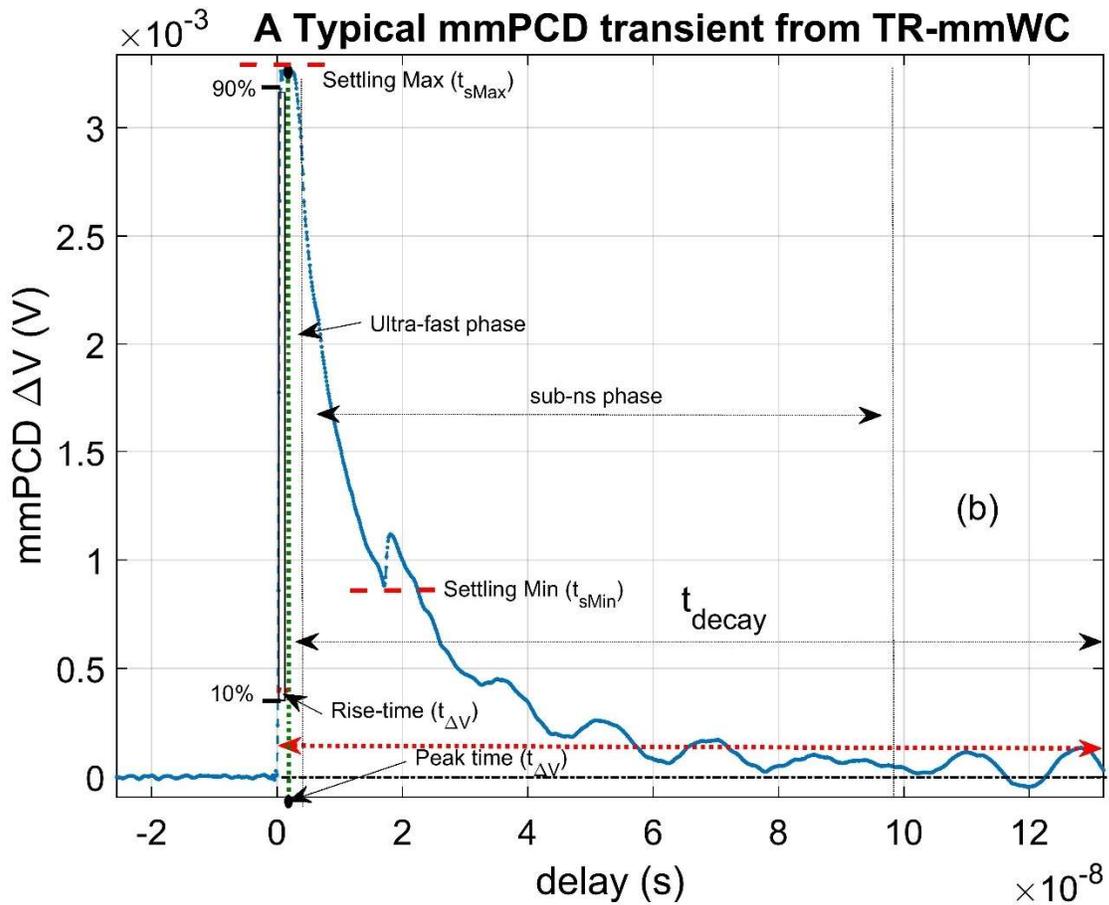

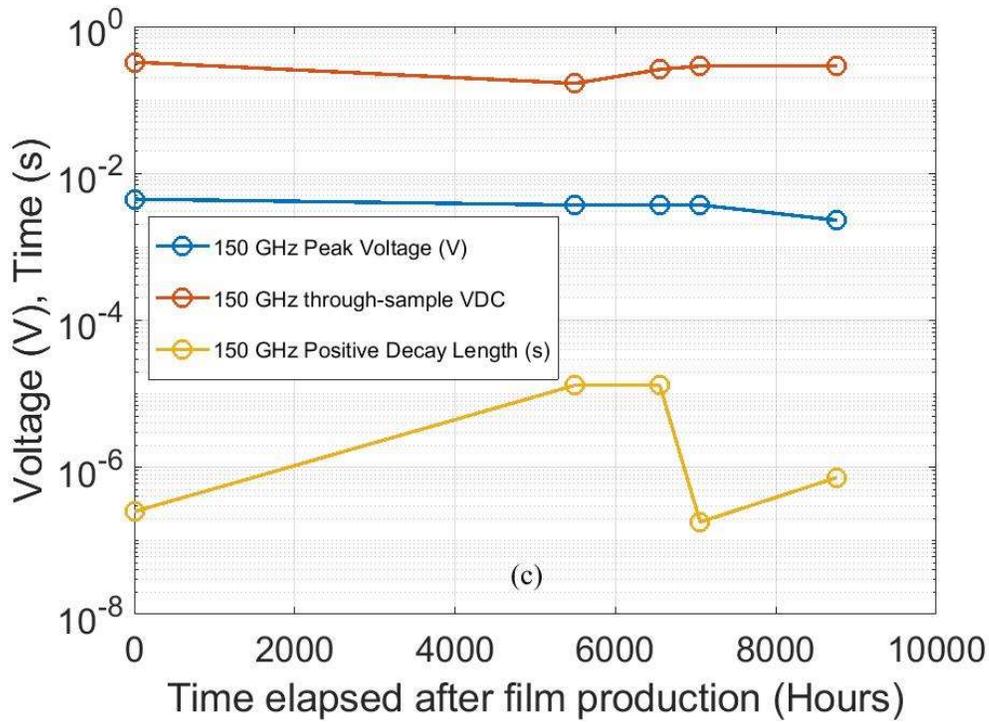

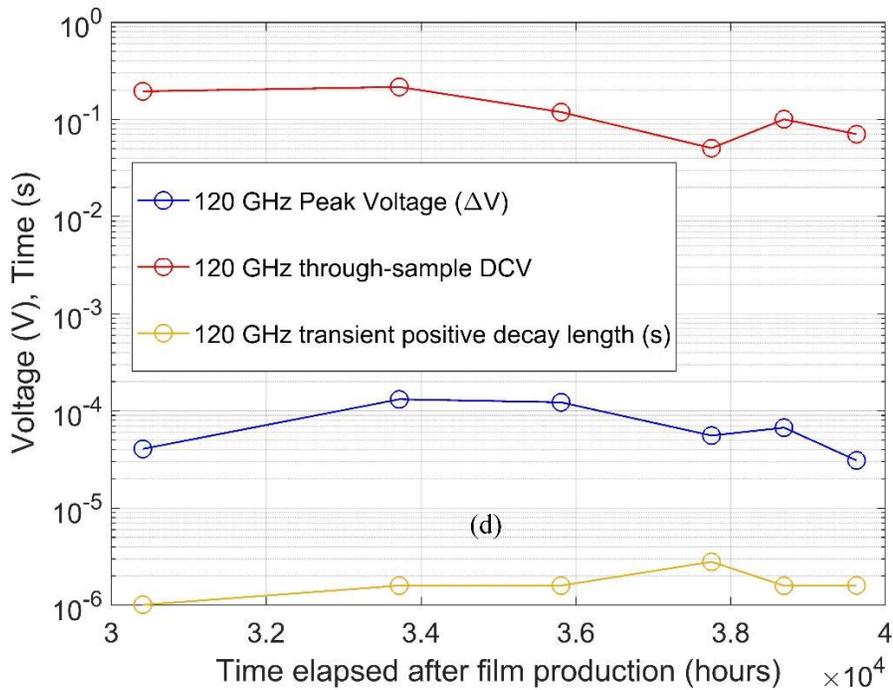

**Figure 1:** (a) Plot of transient signals shown in logarithmic X and Y scales. These raw data are smoothed for 150, and 120 GHz respectively when the encapsulated perovskite sample is placed in the path of probe transmission and illuminated by a 0.69 ps pulse-width 532 nm laser with a fluence of 10.7 µJ cm$^{-2}$. The aging kinetic trace is shown by color scheme, an asterisk (*) representing the 150 GHz data, and colored dots (•) representing 120 GHz probe data that are averaged 4096 times each. (b) shows a typical mmPCD kinetic trace that is stored for each day after averaging 4,096 times. various, time-, voltage features based on step response used for GPR-based prediction whose details are in sec. 3.3.1; (c) the peak AC voltages, DC through voltages, and the positive transient decay length in seconds are shown as acquired at 150 GHz, and (d) the same as (c) but, for 120 GHz probe frequency.

### 3.2 Step-responses and changes in transmission coefficient by age of perovskite

Considering the mmPCD signal analogous to a single input, single output (SISO) continuous, time-invariant dynamic system and the size of the response vector ($\Delta V$) is acquired with the same number of delay in seconds after the laser is turned off. The decay begins just following the settling maximum voltage, we plot the 4 important directly acquired step-responses by respective sample-age, to note additional behavior of material property and their effects. The directly obtained step response data considered in GPR model are, the signal 10%-90% rise-time ($t_{rise}$), representing the intermediate carrier concentration states when the sample is excited by the laser before achieving the full level of photoconductance, the settling time ($t_s$) which is analogous to the recombination period of the signal, except for the fact that this is the time it takes to achieve the response reaching 2% of the peak voltage ($\Delta V$) achieved at the peak time. Peak time ($t_{\Delta V}$) in seconds is the only step response that shows a positive correlation with perovskite sample age in hours.

Figures 2(a) and 2(b) show these four-step responses as a function of sample age in hours acquired at 150 GHz and 120 GHz respectively. Two additional datasets are included while plotting the 120 GHz data shown in Fig. 2(b) considering 10/01/2021, and 11/09/2021 data with time elapsed after film production at 38,688 hours and 39,624 hours respectively. It is readily noted that the $t_{rise}$ is the most varying step-response at both the probe frequencies, and, in general, the baseline rise-time at 150 GHz at ages between 0 and 10,000 hours are longer than those obtained at 120 GHz at ages between 30,000 and 40,000 hours.

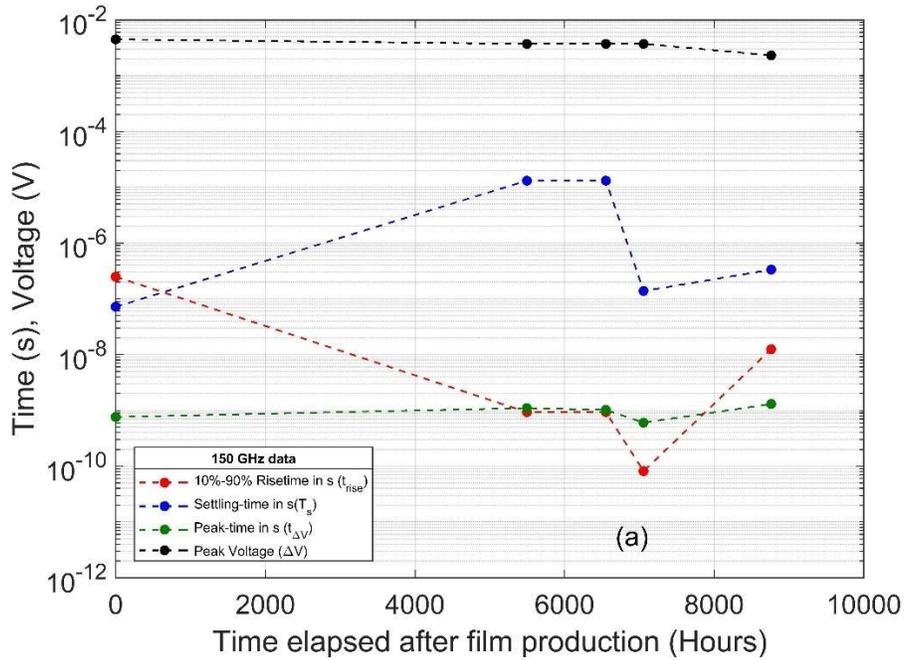

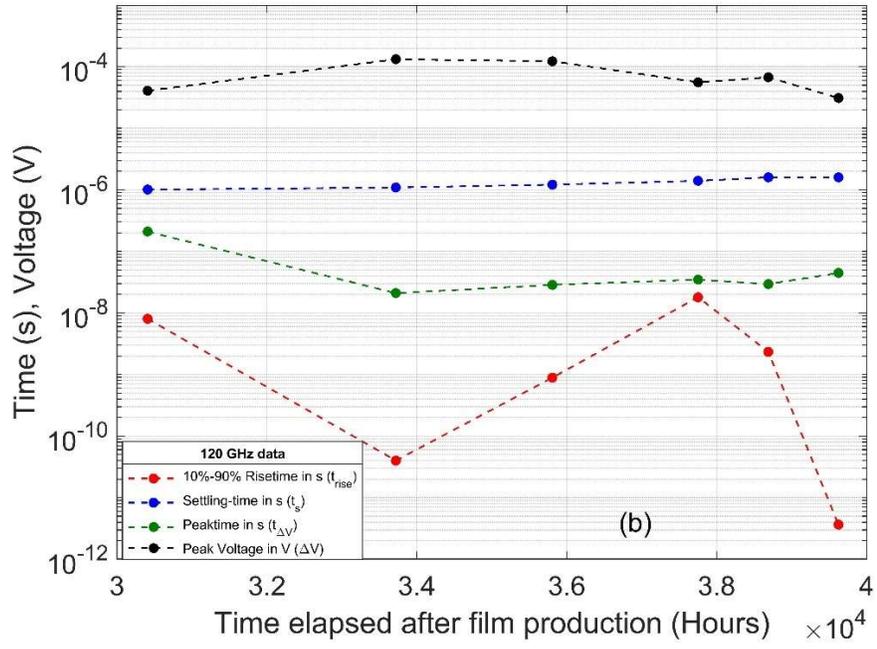

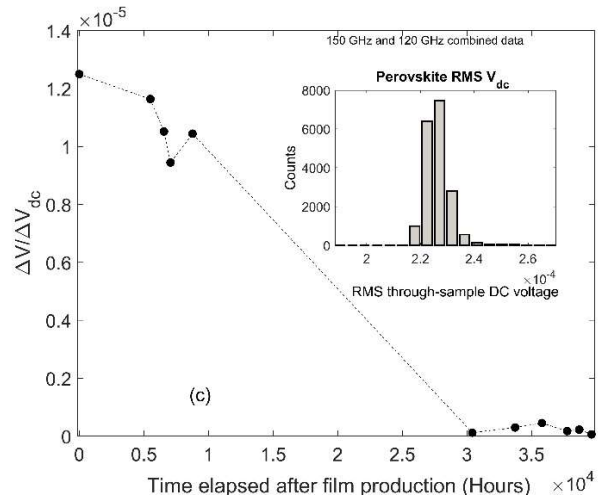

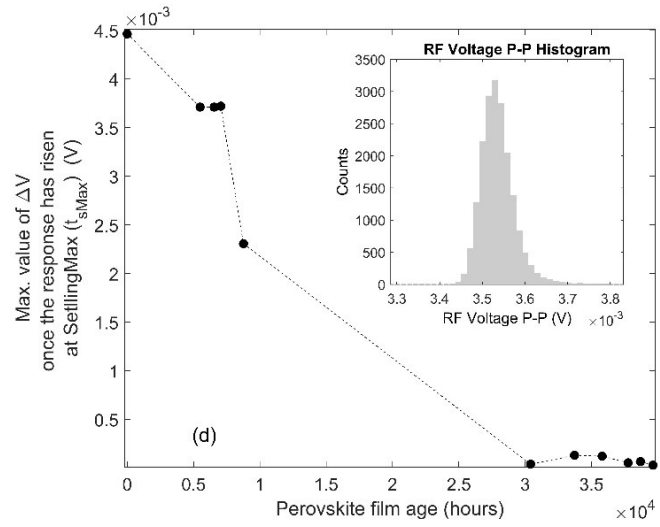

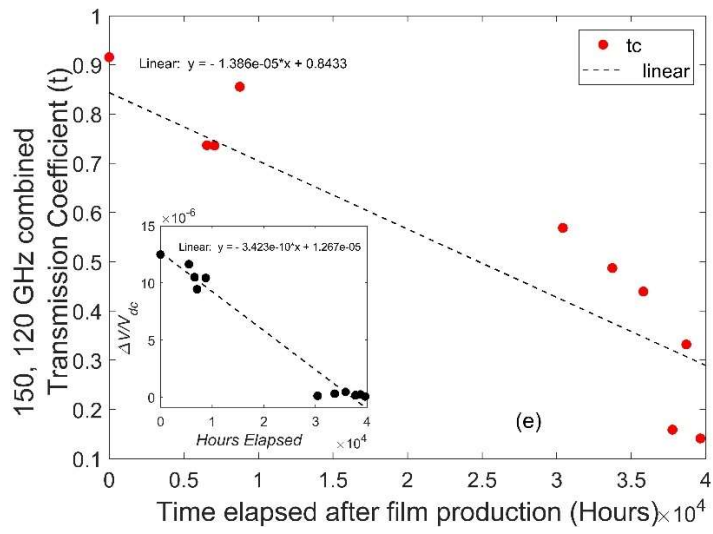

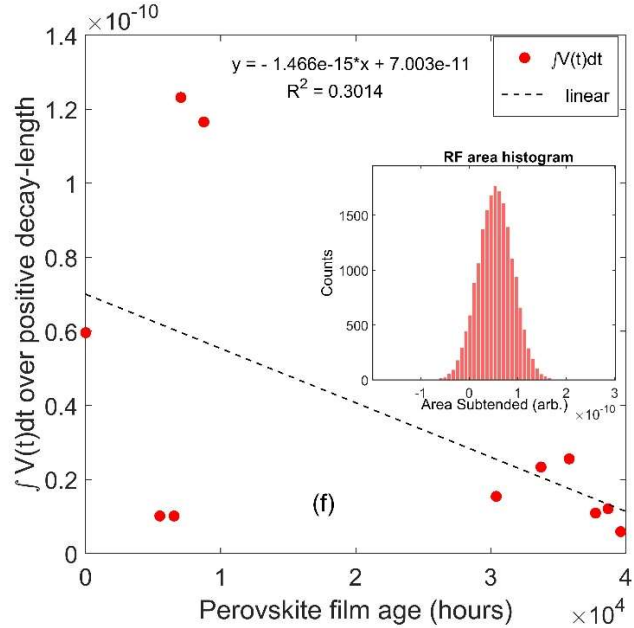

**Figure 2:** (a) shows the 4 step-response parameters taken from the averaged kinetic trace on each date between 0 and 10,000 hours (150GHz); (b) same as in (a) but, when the age of the sample is between 10,000 and 40,000 hours acquired using 120 GHz response; (c) shows the variation of TR-mmWC mmPCD response ratio with sample age (all data combined) with an inset showing the histogram of root mean square value of the DC voltage through sample when the laser is turned off; (d) shows the max value of peak when the response has risen to $t_{sMax}$, with an inset showing the histogram of the peak-to-peak RF voltages collected on one of the session; (e) shows the transmission coefficient of the perovskite film by its age in hours, inset shows the TR-mmWC response ratio $\Delta V/V_{dc}$ as a function of sample age; (f) shows the voltage-time product for each average transient stored for 11 unique periods with an inset that shows the RF area (in Wb.) histogram of 1 day directly from the digital storage oscilloscope (raw data without bias correction), note: the negative area is due to no bias correction applied on the fly.

### 3.3 Supervised learning and prediction of perovskite sample age and its performance
#### 3.3.1 Predictor and response training data for the GPR model

We tested a nonparametric kernel-based probabilistic machine-learning algorithm based on Matlab's Gaussian Process Regression (GPR) learner model. For the training dataset, we identified a set of 12 extracted charge-dynamical and step-response characteristics using the averaged mmPCD transients (kinetic traces). These extracted parameters form the set of predictors. The responses are the age of the sample. A total of 11 different experiments were carried out using probe frequencies either at 150 GHz (between 0 to 9000 hours of the age of the perovskite sample) or at 120 GHz probe frequency (30,000 to 40,000 hours of the age of the sample). The features for GPR model identified are, peak-voltage ($\Delta V$) which has the significance of depicting the maximum absorption of the wave energy due to collisional interactions with charge carrier with density $N_c$, the exact peaking time ($t_{\Delta V}$), the signal-extent, $t_{decay}$ (the maximum period at which the voltage decays to zero, or, the decay time), the TR-mmWC mmPCD response ($\Delta V/V_{dc}$), $V_{dc}$ being the DC voltage registered when the laser beam is cut off and microwave is passing through the sample in transmission mode as measured by the antenna fed Schottky detector [25]. The response ratio determines the photo-conductance in the sample and is proportional to the charge carrier (electron and hole) mobility ($\mu$), $N_c$, photon flux (F), and the system calibration constant K as explained in previous section. Transmission coefficient (tc) being the ratio of the DC voltage due to passage of microwave through sample when the laser is turned off (dark) to the DC voltage directly measured when the sample is removed from the microwave probe beam path. The tc is calculated after taking into account of the reflection loss off the sample surface (~8.3%). This can quantitatively contribute to the estimate of the microwave power dissipation inside the sample arising due to absorption by carriers at room temperature, scattering, and the transformation to immobile charge oscillations in the medium, and tc is a measure that can be measured and connected to the calculation of the sample's dielectric constant, ε [26]. The area under the positive curve $\int V(t)dt$, is an indicator that corresponds to a measure with a unit of magnetic-flux, which is analogous to the fact that the sample itself acts as a capacitor, and the photocurrent so developed after laser illumination effectively discharges it, thereby, the RF area falls off completely over the $t_{decay}$ period. In presence of an electric field over fomamidinium containing perovskite material a few authors have attributed observing the vibrational Stark effect [27] and similar while explaining the photophysics of organic material PSF-BT mentioned as dynamic Stark effect [28]. In our case, the photogenerated charge carriers in the hybrid perovskite are exposed to the RF electric field from the 150/120 GHz probe beam. There is a possibility of vibrational Stark effect happening as we probe the sample even at these incredibly low electric fields. The step-response characteristics of the dynamic system such as the $t_{rise}$, the settling-time ($t_s$) time for the error e(t)=|V(t)-$V_{final}$| between the response ΔV(t) and the steady state response $V_{final}$ to fall below 2% of the peak value, the settling minimum ($t_{sMin}$) is the minimum value of ΔV(t) once the response has risen, settling max ($t_{sMax}$) is the maximum value of ΔV(t) once the response has completely risen (analogous to the root-mean-square value), overshoot ($O_{ΔV}$) is the percentage overshoot relative to the final value of ΔV, and, undershoot ($U_{ΔV}$), the percentage of the undershooting voltage of the averaged signal. Table 2 below provide the Pearson's correlation coefficient of these transient parameters with sample-age in hours.

**Table 2:** Pearson's correlation coefficients of the 5 signal parameters (highlighted in bold) directly related to the charge dynamics in the sample (with a *) and other 7 signal step response-based features of the kinetic traces. Significances of the time- or voltage response from the signals are mentioned for a few of the features selected for the GPR-based age prediction and analysis.

|  | ΔV* | $t_{ΔV}$ | $t_{decay}$* | ΔV/DC* | $t_{rise}$* | $t_s$ | $t_{sMin}$ | peak at $t_{sMax}$ | $O_{ΔV}$ | $U_{ΔV}$ | tc* | $\int V(t)dt$ |
|---|---|---|---|---|---|---|---|---|---|---|---|---|
| Correlation with sample age in hours | **-0.97** | 0.42 | **-0.37** | **-0.98** | **-0.44** | -0.39 | 0.17 | **-0.97** | 0.23 | 0.25 | **-0.86** | **-0.55** |
| Significance | $N_c$ | e-hole recomb. period | trapping/defect density, radiative efficiency | μ, K, $N_c$, F | analogous to 3dB bandwidth |  |  | peak conductance indicator |  |  | Power dissipation, scattering, and evanescence, ε | Voltage-time product, in Webers |

Our selection of a total of 5 out of 12 sample properties and step response parameters is based on the negative correlation.

### 3.3.2 The GPR technique

We are dealing with developing the prediction of a well-encapsulated perovskite sample age by using a high-dimensional extracted photo response transient data. The GPR technique is essentially a nonparametric technique of data fitting to a set of noisy training data that have weak or no (normal) distribution. Hence, it is an ideal technique for supervised learning. Considering a training set {($x_i$, $y_i$); i=1,2….n}, The Matlab-based GPR technique is used to predict the response $y^*$ for a given $x^*$ input vector satisfying a linear regression with an added Gaussian noise $Z_i$ model represented as

$$y_i = x_i^T \gamma + Z_i \qquad (1)$$

Where, superscript T is the transpose of vector x, γ are the coefficients that are estimated from training data, and the Gaussian noise $Z_i \approx N(0, σ^2)$, the stochastic noise (N) with zero mean, $σ^2$ being the error variance. $Z_i$ is independently distributed. This is where a Gaussian Process (GP) is introduced. The GPR introduces latent variables (predictors) f($x_i$), i=1,2,3….n, using a GP and introducing the explicit basis functions 'h'. The basis function 'h' projects the predictor inputs x into a p-dimensional feature space. In our case, the 11 data points we have for each variable mentioned in Table 2 (the averaged transient signal parameters) are random. This random variable set is expected to have a joint Gaussian distribution, and, hence, the joint distribution of the RVs such as $f(x_1)$, $f(x_2)$,… $f(x_n)$ is also Gaussian. It is well known that the GP is associated with a mean function and a covariance function. The covariance function $k(x, x^*)$ captures the smoothness of the response and the mean m(x) is the expectation value E(f(x)) considering $x^*$ as the query point and x being the original set of sample age data, If we now consider the response y, taking into account of the explicit basis function and the Gaussian distributions f(x), we may re-write the prediction model,

$$y = h(x)^T \gamma + f(x) \qquad (2)$$

Whereby, the Gaussian noise $Z_i$ is replaced by the Gaussian function $f(x) \sim GP(0, k(x,x^*))$ which is zero mean GP with covariance (kernel) function $k(x,x^*)$. $\gamma$ is the p x 1 vector of the basis function coefficients. Using this, we can transform the original feature vector (x) into a new feature vector h(x).
The covariance function provides the covariance of training data-based RV to a predicted space using a query value. Considering Gaussian functions f, and $f^*$ corresponding to observations and query points x and $x^*$ respectively, it can be pointed out that the joint distributions of f, $f^*$ and x, $x^*$ also bear a joint distribution by themselves,

$$\begin{bmatrix} f \\ f^* \end{bmatrix} \sim N \left( 0, \begin{matrix} k(x,x) & k(x,x^*) \\ k(x^*,x) & k(x^*,x^*) \end{matrix} \right) \qquad (3)$$

Based on the representation in Eq. (3) we can denote the bivariate conditional probability [29]

$$f^*|f \sim N(k(x^*,x)k(x,x)^{-1}f, k(x^*,x^*) - k(x^*,x)k(x,x)^{-1}k(x,x^*)) \qquad (4)$$

Using Eq. (4) the best estimate of $f^*$ is the mean defined which is, $(k(x^*,x)k(x,x)^{-1}f)$ and the uncertainty in this estimate is the variance term in Eq. (4) which is, $k(x^*,x^*) - k(x^*,x)k(x,x)^{-1}k(x,x^*)$. The GPR-based thin film sample age prediction routine then uses the joint Gaussian distribution of the learning data and uses the covariance matrix so developed for application to the bivariate conditional probability as discussed.

Based on the mmPCD signal, the average parameter step responses variation by age area already discussed in Sec. 3.1 and 3.2 respectively. The 12 predictor values $x_i$ is expected to have very close response values $y_i$. The covariance function captures the smoothness of the similarity in the target values $y_i$ the final selection of the parameters used is only 5 out of 12 due to its strong negative correlation coefficient with the age of the sample itself.

A Gaussian function is completely represented by its mean function m(x) and the covariance function $k(x, x^*)$. Equation (4) designates the relationship between the predicted age from the joint Gaussian distribution

$$f^*|x^*, x, f \sim GPR(m(x), k(x,x^*)) \qquad (5)$$

Covariance kernels functions $k(x, x^*)$ used in the analysis are built in the MatLab regression learner toolbox. The Kernel (covariance) functions include the rational quadratic, squared exponential, Matern-5/2, and exponential GPR. These kernels are well established for use with the GPR-based machine learning model and are well explained in [30]. We have presented some of the model training results after running the rational quadratic, the squared exponential, the Matern-5/2, and the exponential GPR kernels respectively using the MatLab regression learner's toolbox.

### 3.3.2 The GPR model results

Our objective was to demonstrate how the GPR model performs after training and testing on the same age-dependent high-dimensional dataset. We ran the GPR prediction model using all 4 different kernel options once after taking the 5-fold cross-validation (X), a resampling technique, and another time using the same kernel with no X-validation option. The x-validation technique protects the model against overfitting by partitioning the dataset into folds and estimating the accuracy of each fold discretely. Since we used 5-fold X-validation, it means we partitioned 20% of data in each fold and the rest of the data are used for self-testing. In both types of runs, we did not consider optimization options i.e., the hyperparameter options are disabled. With no X-validation option, the prediction accuracy improves substantially, but the fittings become doubtful, especially with this high dimensional training dataset working with the rational quadratic and squared-exponential kernels in the GPR model. They spew out similar predictions. Hence, we compare the Matern-5/2 and exponential kernel-based predictions only once taking the 5-fold X-validation and another time with no X-validation applied rather not protect against overfitting of the model. The results of all the runs are compiled in Table 3

**Table 3:** Gives the trained model derived magnitudes of regression accuracies* in the same units of measurement vector (hours) such as the root mean square error (RMSE), Mean Squared Error (MSE), Mean Absolute Error (MAE), the $R^2$ value and the training periods in seconds. Results obtained for model runs with 4 different kernels with a 5-fold X-validation option and no X-validation option.

| | RMSE (h) | | $R^2$ | | MSE (h) | | MAE (h) | | Training Time (s) | |
|---|---|---|---|---|---|---|---|---|---|---|
| **Kernel used** | *5-fold X-Val.* | *No X-Val.* | *5-fold* | *No* | *5-fold* | *No* | *5-fold* | *No* | *5-fold* | *No* |
| Rational Quadratic | 3030.8 | 536.72 | 0.97 | 1.0 | 9.18e+06 | 2.88e+05 | 2252.4 | 436.5 | 1.08 | 0.29 |
| Squared Exponential | 3032.3 | 536.72 | 0.97 | 1.0 | 9.19e+06 | 2.88E+05 | 2260.4 | 436.46 | 0.23 | 0.04 |
| Matern-5/2 | 2780.7 | 490.09 | 0.98 | 1.0 | 7.73e+06 | 2.40e+05 | 2169.3 | 490.0 | 0.19 | 0.05 |
| Exponential | 3496.6 | 7.017 | 0.97 | 1.0 | 1.22e+7 | 49.24 | 2493.6 | 4.92 | 0.20 | 0.05 |

*$RMSE = \sqrt{\frac{1}{n}\sum |y_i - \hat{y}|^2}$, $MSE = \frac{1}{n}\sum |y_i - \hat{y}|^2$, $MAE = \frac{1}{n}\sum |y_i - \hat{y}|$, $\hat{y}$ is the average value, and $|y_i - \hat{y}|^2$ is the variance-dependent total sum of squares (SS$_{tot}$), and, the coefficient of determination (COD), $R^2 = 1 - \frac{SS_{res}}{SS_{tot}}$, SS$_{res}$ is the residual sum of squares; residual values are shown as insets to Figs. 3 below.

It is noted from Table 3 that the use of rational quadratic, and the squared exponential kernels both yield the same results however, the use of Matern-5/2 and exponential kernels yield more realistic predicted values, both, with X-validation and without. Comparing the data from Table 3 and Figure 3 below we note that the Matern-5/2-based predictions are more realistic than those obtained using the exponential kernel. The training speeds are also reasonable for Matern-5/2. In general, the COD being 1.0 itself in all the runs using no X-validation option and prediction space exactly matching with the data itself is a bit doubtful outcome. We need to test the already-built GPR model in the training process using a long set of test data for continuous prediction which will be our next step. Since our experimental data sample space is very small, we did not feel the necessity to use principal component analysis (PCA) in the algorithm to aid in dimensionality reduction, hence, in all model runs, the PCA option was turned off.

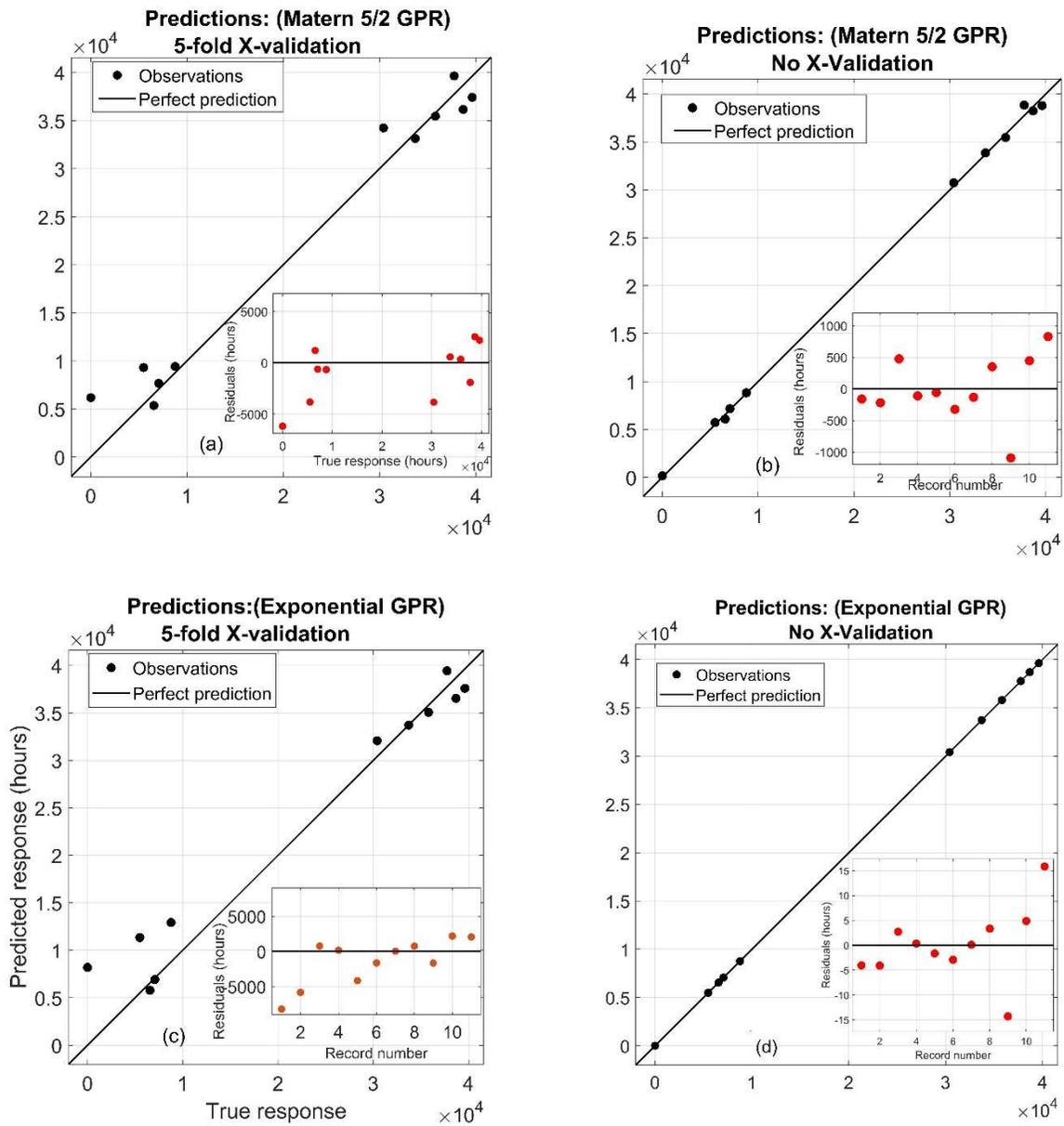

**Figure 3:** shows the predicted and actual plot using the GPR exponential kernel, (b), (c), and (d) show the same as (a) but when using the rational-quadratic, squared-exponential, and Matern-5/2 kernels, (e) shows the residual plot for the exponential GPR which is best adopted using the 5 selected mmW transient predictors.

It is readily noted from Fig 3 that the exponential GPR with No X-Val. (Figure 1d) seems to yield an overly trained prediction output as the predicted response matches 100% with the true response. This might be a case of noise capture resulting in a deviation of the GPR approximation to produce a non-generalized model run. Neglecting 1(d) and all no X-validation outputs with COD=1.0, from the results we consider only the 2 runs that best yield the predicted results after training and testing on the same aging sample millimeter-wave response data. The first candidate to use would be the Matern-5/2 run with 5-fold X-Validation in place, and the second candidate would be the exponential GPR kernel derived run outputs. The exponential kernel runs with or without X-validation and provides incredibly accurate outputs which might signify over-fitting in place. However, Both Matern-5/2 and exponential kernels provide a very reasonable COD value, and the prediction values scatter from the true responses quite tightly. Considering Matern-5/2 kernel to be more realistic, we plotted the GPR model responses with Matern-5/2 kernel and plotted by the record numbers 1-11 in figures 4(a) and 4(b) respectively for once account the 5-fold X-validation, and another time not accounting any resampling technique. We note that the latter method once again yields a very accurate prediction. Hence, we have to test the models with an independent test data set to be able to ascertain the final model performance. For the present moment, we will consider Matern-5/2 kernel more rationally suitable to our application than the exponential kernel-based GPR.

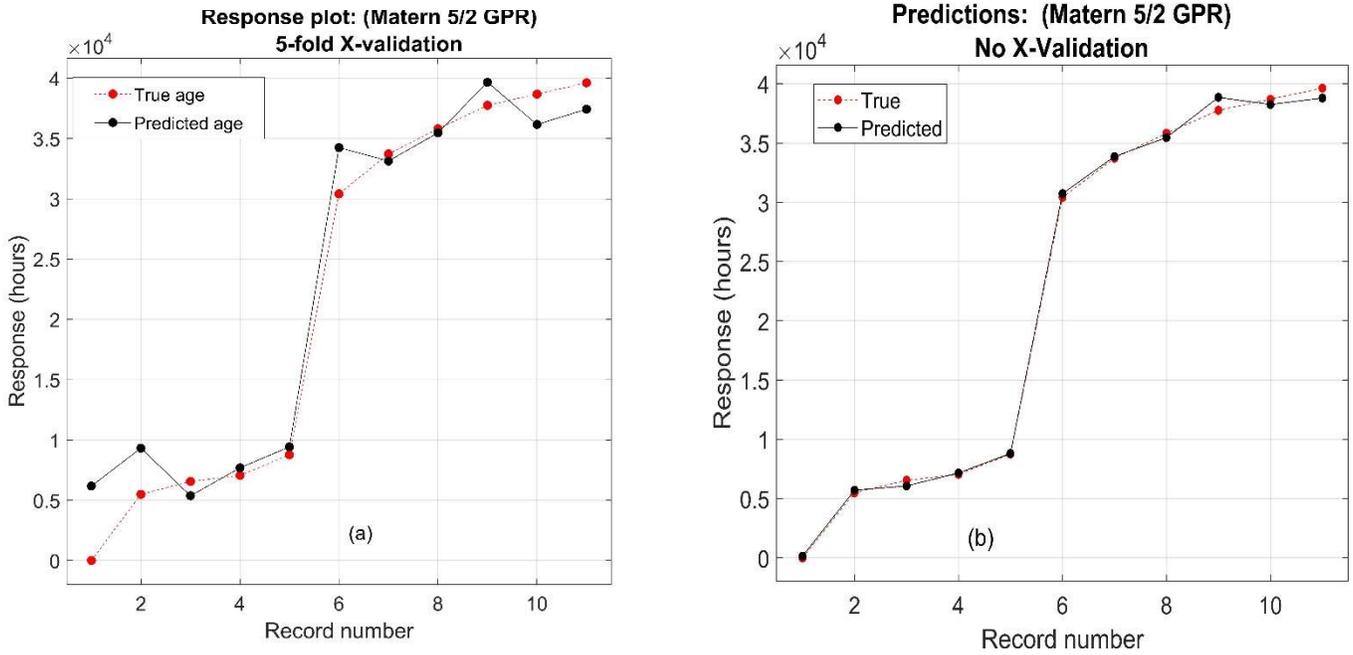

**Figure 4:** Predicted responses as a function of true responses for the Matern-5/2 kernel and exponential GPR kernels once using the 5-fold X-validation (a, and c respectively) and another time with no X-val. Used (b, and d respectively). The insets show the residuals for the sample age in hours elapsed from its manufacturing (in Fig. 3(a)) and, in record number (1-11) in other figures.

## 4.0 Conclusions

For the first time, we have applied a Gaussian Process Regression (GPR) based supervised learning technique using mmPCD signal features from a transmission-mode free space operating TR-mmWC apparatus. The signal parameters are directly used for predicting and testing the age of a well-encapsulated perovskite (MAL3I) thin film sample based on the obtention of negatively correlated mmPCD derived descriptors collected over 0 to 40,000 hours after the film production. An assumption is made that the 120 GHz and 150 GHz mmPCD/optical responses of the sample are comparable and can be combined for generating a total sample space of 11 days of observation. The GPR technique is applied to 5 selected mmPCD features that bear a strong negative correlation with the age of the sample itself, thereby attempting to qualify a sample property degeneracy over time assumption. Out of the 5 selected mmPCD features, 3 are conductance parameters and 2 of them are step responses acquired directly from the averaged transients collected over time. Due to the high dimensionality of the data (total number of features greater than the number of observations), we have been able to choose the GPR technique for sample age prediction based on publication [20, 22]. The rationale behind training a GPR using photoconductance signal-based predictors is to gain faith in a numerical technique aimed at directly classifying the material age based on microwave absorption-based responses alone. This might aid in the process of automating the reporting of the aging semiconductor component used in the larger format device. This study has also revealed passively, the direct relationship of some of the signal step responses of the sample-derived averaged transients (generated due to differential absorption of the probe beam through the sample when illuminated momentarily by a 532 nm laser with moderate fluence), and some charge dynamical features such as the peak voltage (carrier concentration), the $\Delta V/V_{dc}$ (photo-conductance parameter), the mmW transmission coefficient which is a dielectric response of the sample that have been used to generate DC responses at 150 GHz and 120 GHz respectively.

Broadly, we conclude the following based on this study:

- The differential absorption-based mmPCD signals acquired using the TR-mmWC system at 120 and 150 GHz show degeneracy with time resulting in a strong negative correlation with the age of the sample.
- The GPR model building routine as a resident in MatLab [30] can be suitably used for the high dimensional dataset such as the one presented here for predicting the sample age by testing on the same data as training and following the recommendation in a paper by Zhang et al. (2018) [22].
- We introduce the RF voltage-time area (flux-total) parameter in this training sequence which is the integral of the AC voltages of the mmPCD signal acquired over the $t_{decay}$ (positive part of the transient) and albeit, it does not show a very strong negative correlation with the sample age, however, we find that including this feature in training the GPR model improves the prediction.
- Amongst the 4 major GPR kernel-based non-parametric models, we finally could make use of the two kernels viz., the Matern-5/2 and the exponential, that provide reasonable prediction space when compared with the true response. Our study reveals that the Matern-5/2 kernel provides a reasonable prediction of sample age with acceptable statistics. However, the use of the exponential kernel yields prediction values and CODs that seem to be overwhelmingly accurate. The choice of whether to use X-validation to suppress overfitting in the model runs or not remains a challenge.
- We need to collect more independent aging data from the still existing and working MAL3I sample at 120, and 150 GHz and then test the GPR models developed here by splitting the training and testing sets.

In the future, we will also test the algorithm with the sample responses obtained when the MAL3I sample is exposed to a very wide range of temperatures. Using this forward sample age prediction modeling technique it could be possible to follow the rules of inverse modeling [31] to obtain the step response and photoconductance parameters, and charge carrier generation efficiency of the perovskite film for known conditions of the sample, just by using its age in hours. More detailed photophysics of the perovskite sample can be generated by taking the transmission and reflection-based response signals as a function of closely spaced probe frequency and probe power levels, and, by exposing the sample to external electric fields to observe the electro-modulated differential absorption.

**Acknowledgments**
BR acknowledges the STIR grant received from the Army Research Office (ARO) award number W911NF2210116 and NSF for studying the structure-property relationship of perovskite thin films and to National Science Foundation (NSF) for awarding a catalyst grant for studying millimeter wave responses in aging perovskites through Award Number 2200518. MHW acknowledges funds received from NSF-partnership for Research and Education in Materials (PREM) Award Number 1523617. All authors acknowledge the Consortium for Nuclear Security Advanced Manufacturing Enhanced Machine Learning funded by the Department of Energy/National Nuclear Security Administration under Award Number NA0003979. The authors immensely thank Drs. Franky So and Stephen Amoah of N.C. State University for providing the state-of-the-art encapsulated MAL3I sample based on which these experiments were done.

**References**

[1] W.E.I. Sha, X. Ren, L. Chen, and W.C.H. Choy, The efficiency limit of $CH_3NH_3PbI_3$ perovskite solar cells, Appl. Phys. Lett. 106, 221104 (2015). http://dx.doi.org/10.1063/1.4922150.

[2] H. Wei, and J. Huang, Halide perovskites for ionizing radiation detection, Nature Communications, 10:1066 (2019) https://doi.org/10.1038/s41467-019-08981.

[3] I. Maeng, S. Lee, H. Tanaka, J-H Yun, S. Wang, M. Nakamura, Y-K Kwon, and M-C Jung, Unique phonon modes of $CH_3NH_3PbBr_3$ hybrid perovskite film without the influence of defect structures: an attempt toward a novel THz-based application, NPG Asia Materials, 12:53 (2020) https://doi.org/10.1038/s41427-020-0235-6.

[4] M.B. Johnston, and L.M. Herz, Hybrid perovskites for photovoltaics: charge-carrier recombination, diffusion, and radiative efficiencies, Acc. Chem. Res. 49 (2016) 146-154. https://doi.org/10.1021/acs.accounts.5b00411.

[5] D. W. deQuilettes, K. Frohna, D. Emin, T. Kichartz, V. Bulovic, D.S. Ginger, S.D. Straks, Charge-carrier recombination in halide perovskites, Chem. Rev. 20 (2019) 11007-11019. https://doi.org/10.1021/acs.chemrev.9b00169.

[6] Z. Cheng and D.M. O'Carroll, Photon recycling in semiconductor thin films and devices, Adv. Sci. 8 (2021) 2004076. https://doi.org/10.1002/advs.202004076.

[7] H. Oga, A. Saeki, Y. Ogomi, S. Hayase, and S. Seki, Improved understanding of the electronic and energetic landscapes of perovskite solar cells: High local charge carrier mobility, reduced recombination, and extremely shallow traps, J. Am. Chem. Soc., 136 (39) (2014) 13818-13825. https://doi.org/10.1021/ja506936f.

[8] O.G. Reid, M. Yang, N. Kopidakis, K. Zhu, and G. Rubles, Grain-size-limited mobility in methylammonium lead iodided perovskite thin films, ACS Energy Lett. 1 3 (2016) 561-565. https://doi.org/10.1021/acsenergylett.6b00288.

[9] K-S. Lee, R. Kang, B. Son, D-Y. Kim, N.E. Yu & D-K. Ko, All-optical THz wave switching based in $CH_3NH_3PbI_3$ perovskites, Sci. Rep. 6 (2016) 37912. https://doi.org/10.1038/srep37912.

[10] S. Ahmad, A. Husain, M., M. A. Khan, I. Khan, A. Khan, A.M. Asiri, 5-Perovskite-based material for sensor applications, ACS Appl. Mater InterfacesHybrid Perovskite Composite Materials, Design to Applications, Woodhead Publishing Series in Composites Science and Engineering, 135-145 (2021). https://doi.org/10/1016/B978-0-12-819977-0.00005-6.

[11] A. Sultana, S.K. Ghosh, M.M. Alam, P. Sadhukhan, K. Roy, M. Xie, C.R. Boweb, S. Sarkar, S. Das, T.R. Midyya, D. Mandal, Methylammonium lead iodide incorporated Poly(vinylidene fluoride) nanofibers for flexible piezoelectric-pyroelectric nanogenerator, ACS Appl. Mater. Interfaces Vol. 11, 30 27279-27287 (2019). https://doi.org/10.1021/acsami.9b04812.

[12] D. Baranov, A. Fieramosca, R. X. Yang, L. Polimeno, G. Lerario, S. Toso, C. Giansante, M. D. Giorgi, L.Z. Tan, D. Sanvitto, and L. Manna, Aging of self-assembled lead halide perovskite nanocrystal superlattices: Effects on photoluminescence and energy transfer, ACS Nano 15 (2021) 650-664. https://doi.org/10.1021/acsnano.0c06595.

[13] B. Roy, C.R. Jones, B. Vlahovic, H. W. Ade, and M. H. Wu, A time-resolved millimeter wave conductivity (TR-mmWC) apparatus for charge dynamical properties of semiconductors, Rev. Sci. Instrum. 89 (2018) 104704. https://doi.org/10.10631/1.5026848.

[14] J. Schmidt, M. R. G. Marques, S. Botti, and M. A. L. Marques, Recent advances, and applications of machine learning in solid-state materials science, npj Computational Materials 5:83 (2019) 36 pages. https://doi.org/10.1038/s41524-019-0221-0.

[15] N. Wagner, and J. M. Rondinelli, Theory-guided machine learning in materials science, Front. Mater. 3:28 (2016) 9 pages. https://doi.org/10.3389/fmats.2016.00028.

[16] R. Ramprasad, R. Batra, G. Pilania, A. Mannodi-Kanakkithodi, and C. Kim, Machine learning in materials informatics: recent application and prospects, npj Computational Materials, 3:54 (2017) 13 pages. https://doi.org/10.1038/s41524-017-0056-5.

[17] J.L. Lansford, and D.G. Vlachos, Infrared spectroscopy data- and physics-driven machine learning for characterizing surface microstructures of complex materials, Nature Communications 11:1513 (2020) 12 pages. https://doi.org/10.1038/s4167-020-15340-7.

[18] A. Seko, T. Maekawa, K. Tsuda, and I. Tanaka, Machine learning with systematic density-functional theory calculations: Applications to melting temperatures of single and binary component solids, ArXiv:1310.1546v1 ]cond-mat.mtrl-sci] (2013) 9 pages.

[19] K.T. Butler, J. M. Frost, J. M. Skelton, K. L. Svane, and A. Walsh, Computational materials design of crystalline solids, Chem. Soc. Rev. 45 (2016) 6138-6146. https://doi.org/10.1039/c5cs00841g.

[20] Q. Tao, P. Xu, M. Li, and W. Lu, Machine learning for perovskite materials design and discovery, npj Computational Materials 7:23 (2021) https://doi.org/10.1038/s41524-021-00495-8.

[21] L. Zhang, M. He, and S. Shao, Machine learning for halide perovskite materials, Nano Energy 78 (2020) 105380 https://doi.org/10.1016/j.nanoen.2020.105380.

[22] N. Zhang, J. Xiong, K. Leatham, Gaussian process regression method for classification for high-dimensional data with limited samples, 2018 Eight International Conference on Information Science and Technology (ICIST), 30 June 2018-06July 2018, Cordoba, Granada, and Seville, Spain (2018) https://doi.org/10.1109/ICIST.2018.8426077.

[23] D. Zhao, M. Sexton, H-Y Park, G. Baure, J.C. Nino, and F. So, High-efficiency solution processed planar perovskite solar cells with a polymer hole transport layer, Adv. Energy Mater. 5 (2015) 1401855 https://doi.org/10.1002/aenm.201401855.

[24] Y. Luo, S. Liu, N. Barange, L. Wang, and F. So, Perovskite solar cells on corrugated subtrates with enhanced efficiency, Small 12 No. 46 (2016) 6346-6352 https://doi.org/10.1002/smll.201601974.

[25] J. L. Hesler and T. W. Crowe, Responsivity and noise measurements of zero-bias Schottky diode detectors, 18[th] International Conference on Space Terahertz Technology, (ISSTT 2007), 11-18 May 2007, Pasadena, CA USA (2007) 89-92.

[26] G. Kozlov, and A. Volkov, Coherent source submillimeter wave spectroscopy, In: G. Grüner (Ed.), Millimeter and Submillimeter Wave Spectroscopy of Solids, Springer, Berlin, 1998, pp 51-107.

[27] R. Rakowski, W. Fisher, J. Calbo, M. Z. Mokhtar, X. Liang, D. Ding, J.M. Frost, A. A. Haque, A. Walsh, P. R. F. Barnes, J. Nelson, and J. J. van Thor, Nanomaterials 12 (2022) 1616. https://doi.org/10.3390/nano12101616.

[28] V. Gulbinas, R. Kananavičius, L. Valkunas, and H. Bässler, Dynamic Stark effect as a probe of the evolution of geminate electron-hole pairs in a conjugated polymer, Phy. Rev. B 66 (2002) 233203 https://prb/pdf/10.1103/PhysRevB.66.233203.

[29] Gaussian Processes for Regression: A Quick Introduction. GPtutorial.pdf (mebden.com), 2008.

[30] C.E. Rasmussen, and C. K. I. Williams, Gaussian Processes for Machine Learning. MIT Press Cambridge, MA, U.S.A. (2006) https://doi.org/10.7551/mitpress/3206.001.0001.

[31] J-W Lee, W. B. Park, B. D. Lee, S. Kim, N. H. Goo, and K-S Sohn, Dirty engineering data-driven inverse prediction machine learning model, Sci. Rep. 10 (2020) 20443. https://doi.org/s41598-020-77575-0.